\documentclass[11pt]{article}
\usepackage{amsmath,amsfonts,amssymb,amsthm}

\newcommand{\field}[1]{\mathbb{#1}}
\newcommand{\N}{\field{N}}
\newcommand{\Z}{\field{Z}}
\newcommand{\R}{\field{R}}
\newcommand{\C}{\field{C}}

\newcommand{\F}{\mathcal{F}}
\newcommand{\Fin}{\mathcal{F}_{\rm fin}}
\renewcommand{\H}{\mathcal{H}}
\newcommand{\Hel}{\mathcal{H}_{\text{at}}}
\newcommand{\h}{\mathfrak{h}}
\renewcommand{\L}{\mathbf{B}}
\newcommand{\HxF}{\tilde{\mathcal{H}}}   

\newcommand{\vac}{|\text{vac}\rangle}

\newcommand{\Hamel}{H_{\text{at}}}
\newcommand{\Sel}{\Sigma_{\text{at}}}
\newcommand{\SelN}[1]{\Sigma_{\text{at},#1}}
\newcommand{\HamNel}[1]{H_{\text{at},#1}}          
\newcommand{\Hint}{H_{\text{int}}}

\newcommand{\Hex}{\tilde{H}}

\newcommand{\dGamma}{\mathrm{d}\Gamma}

\newcommand{\eps}{\varepsilon}
\newcommand{\ph}{\varphi}
\newcommand{\const}{\mathrm{const}}
\newcommand{\ran}{\mathrm{Ran}}

\newcommand{\restricted}{|\grave{}\,}

\newcommand{\sprod}[2]{\mbox{$\langle #1,#2 \rangle$}}       

\newcommand{\Ran}{\operatorname{Ran}}
\newcommand{\supp}{\operatorname{supp}}
\newcommand{\Ima}{\operatorname{Im}}

\begin{document}
\title{Non-relativistic Matter and Quantized Radiation}
\footnotetext[1]{Supported in part by U.S. National Science
Foundation grant DMS 01-00160.}
\author{\vspace{5pt} M.~Griesemer\footnotemark\\
\vspace{-4pt}\small{Department of Mathematics, University
of Alabama at Birmingham,} \\
\small{Birmingham, AL 35294, USA. {\tt marcel@math.uab.edu}}\\
}
\date{}
\maketitle

\begin{abstract}
This is a didactic review of spectral and dynamical properties of
atoms and molecules at energies below the ionization threshold,
the focus being on recent work in which the author was involved.
As far as possible, the results are described using a simple model
with \emph{one} electron only, and with \emph{scalar} bosons. The
main ideas are explained but no complete proofs are given. The
full-fledged standard model of non-relativistic QED and various of
its aspects are described in the appendix.
\end{abstract}

\section{Introduction}
\label{hbl-sec:intro}

An atom or molecule in an excited state with energy below the
ionization threshold will eventually relax to its ground state by
dissipating excess energy in the form of radiation. This process
of \emph{relaxation to the ground state} is one of the basic
phenomena responsible for the production of all visible light. It
involves a range of energies within a few electron volts; a scale
where the electron-positron pair creation and the production of
ultraviolet radiation is highly suppressed. In a first
mathematical study of relaxation to the ground state it is
therefore reasonable and legitimate to work with a model where the
electron-positron pair creation is entirely neglected and an
ultraviolet cutoff is imposed on the electron-photon interaction.
These simplifying assumptions lead to a mathematically well
defined model of matter, often called standard-model of
non-relativistic quantum electrodynamics, or Pauli-Fierz model.
Since the numerical predictions of this model are in good
agreement with measurement data it is a viable physical model.
Yet, only little mathematically rigorous work on this model had
been done before the middle of the 1990s, when several groups of
researchers started to investigate various of its aspects. Most
influential, perhaps, were the papers of H\"ubner and Spohn
\cite{HuSp1, HuSp2, HuSp3} on spectral and scattering theory, of
Bach et al.~on spectral analysis ~\cite{BFS1, BFS2, BFS3, BFSS},
of Derezi\'nski and G\'erard on scattering theory \cite{DG1, DG2},
and of Jak{\v{s}}i{\'c} and Pillet on thermal relaxation
\cite{JaPi1, JaPi2, JaPi3, JaPi4}. The present article reviews
recent work on the phenomenon of relaxation to the ground state
for states with total energy below the threshold energy for
ionization. It is guided by papers of Lieb, Loss and Griesemer,
and of Fr\"ohlich, Schlein and Griesemer \cite{GLL, Griesemer2002,
FGS1, FGS2}. The focus is on the existence of an ionization
threshold and the localization of the electrons below this energy,
the existence of a ground state, and the existence and
completeness of many-photon scattering states (asymptotic
completeness of Rayleigh scattering).

The results to be discussed on existence of a ground state and on
the localization of photons with energy below the ionization
threshold, unlike previous results, hold for all values of the
physical parameters such as the fine structure constant and the
ultraviolet cutoff. This is crucial for moving on to physically
more realistic models without ultraviolet regularization. The
analysis of electron-photon scattering is based on methods and
ideas from the scattering theory of $N$-body quantum systems
\cite{SigalSoffer1987, Graf1990, Yafaev1993, GrafSchenker1997}. On
the one hand the electron-photon dynamics is easier to analyze
than the full $N$-body problem since there is no photon-photon
interaction. On the other hand the number of photons is not
constant! In fact, it might even diverge as time $t\to\infty$.
This divergence is avoided by imposing a cutoff on the interaction
between electrons and low-energy photons (infrared-cutoff). It is
one of the main open challenges in the mathematical analysis of
matter interacting with quantized radiation to prove asymptotic
completeness for Rayleigh
scattering without this infrared-cutoff.\\

This article is organized as follows. In Section 2.1 we begin with
the description of a simple, but non-trivial model of matter and
radiation. There is only one electron, besides the static nuclei,
and the radiation is described by scalar bosons.

Sections \ref{hbl-sec:expo}, \ref{hbl-sec:gs},
\ref{hbl-sec:scatex}, and \ref{hbl-sec:ac} describe the main
results of the papers \cite{Griesemer2002, GLL, FGS1} and
\cite{FGS2}, respectively. Sections \ref{hbl-sec:schroed} and
\ref{hbl-sec:physics} summarize mathematical and physical
background, and Sections~\ref{hbl-sec:no-ev} and
\ref{hbl-sec:relax2gs} are devoted to side issues in the
aforementioned papers.

Section \ref{hbl-sec:Nel} outlines the modification of results and
proofs that are necessary to accommodate $N>1$ electrons, and
Section \ref{hbl-sec:open} ends this review with concluding
remarks and a discussion of selected open problems.

There is a self-contained appendix on the standard model of
non-relativistic QED.\\

\noindent \emph{Acknowledgments.} Most of the content of this
review article I learned from my collaborators  J\"urg Fr\"ohlich,
Elliott H. Lieb, Michael Loss, and Benjamin Schlein. I am indebted
to all of them. I thank David Hasler for his careful proofreading.

\section{Matter and Radiation}

All electrons of an atom or molecule are well localized near the
nuclei if the total energy is below the ionization threshold.
Therefore the number of electrons is inessential for the phenomena
to be described mathematically in this section. To simplify
notation and presentation we restrict ourselves to one-electron
systems; the generalization to $N>1$ electrons is described in
Section~\ref{hbl-sec:Nel}.

\subsection{A Simple Mathematical Model}
\label{hbl-sec:model}

The main features of quantum electrodynamics that are responsible
for the phenomena to be studied, are the peculiar form of
interaction between light and matter, through creation and
annihilation of photons, and the fact that photons are massless
relativistic particles. The spin of the electron and the helicity
of the photons do not play an essential role in most of our
analysis. For the purpose of this introduction we therefore
neglect these subtleties and present a caricature of QED which
only retains the aforementioned main features. The full-fledged
standard model is described in the appendix.

We first introduce our models for matter and radiation separately
before describing the composed system and the interaction.

A (pure) state of a quantum particle, henceforth called
\emph{electron}, is described by a normalized vector $\psi\in
L^2(\R^{3},\C)$, $\int_A|\psi(x)|^2dx$ being the probability to
find the particle in the region $A\subset\R^3$. Its time evolution
is generated by a Schr\"odinger operator
\begin{equation}\label{hbl:schroed}
  \Hamel = -\Delta + V
\end{equation}
where $-\Delta$ is the positive Laplacian and $V$ is the operator
of multiplication with a real-valued function $V(x)$, \(x\in
\R^3\). We assume that $V\in L^2_{\rm loc}(\R^3)$ and that there
exist constants $\alpha<1$ and $\beta$ such that
\begin{equation}\label{hbl:V1}
  \sprod{\ph}{V_{-}\ph} \leq
  \alpha\sprod{\ph}{(-\Delta)\ph}+\beta\sprod{\ph}{\ph}
\end{equation}
for all $\ph\in C_0^{\infty}(\R^3)$, where $V_{-}=\max(-V,0)$.
Hence the operator $-\Delta+V$ is symmetric and bounded from
below, which allows us to define a self-adjoint Hamiltonian
$\Hamel$ by the Friedrichs' extension of $-\Delta+V$.

We shall be most interested in the case where
\begin{equation}\label{hbl:coulomb}
  V(x)=V_{Z}(x) := - \sum_{j=1}^K \frac{Z_j}{|x-R_j|},
\end{equation}
$Z_j,\ j=1\dots K$, are positive integers, and $R_j\in\R^3$. The
function \eqref{hbl:coulomb} is the potential energy (or the
scalar potential in Coulomb gauge) of one electron at $x\in \R^3$
in the field of $K$ nuclei with positions \(R_1,\dots, R_K\) and
atomic numbers $Z_1,\ldots, Z_K$.

The Hamiltonian \eqref{hbl:schroed} with $V$ given by
\eqref{hbl:coulomb} describes a molecule with one electron and
static nuclei in units where the unit of length is
$\hbar^2/(2me^2)=r_B/2$ and the unit of energy is $2e^2/r_B=4$ Ry
(see Appendix~\ref{sec:qed}). Here $r_B=\hbar^2/(me^2)$ is the
Bohr radius, $-e$ is the charge of the electron and $m$ is its
mass. This Hamiltonian is self-adjoint with domain
\(D(\Hamel)=H^2(\R^{3})\), the Sobolev space of twice weakly
differentiable $L^2$-functions \cite{Kato1951, ReedSimon2}.

A pure state of the radiation field is described by a normalized
vector in the \emph{bosonic Fock space over} $L^2(\R^3)$. This is
the space
\begin{equation*}
  \F = \bigoplus_{n\geq 0} \mathcal{S}_n L^2(\R^{3n};\C)
\end{equation*}
where $\mathcal{S}_0 L^2(\R^0):=\C$, and $\mathcal{S}_n$ denotes
the orthogonal projection onto the subspace of square integrable
functions $f(k_1,\dots, k_n)$ that are symmetric with respect to
permutations of the $n$ arguments \(k_1,\dots,k_n\in \R^3\). Such
a function describes a state of $n$ bosons, henceforth called
\emph{photons}, with wave vectors $k_1,\dots,k_n$. The vector
\(\vac=(1,0,0,\ldots)\in \F\) is called the \emph{vacuum} vector.
With $\Fin$ we denote the subspace of sequences $\ph=(\ph)_{n\geq
0}\in\F$ with $\ph_n=0$ for all but finitely many $n\in\N$.

The energy of a state \(\ph=(\ph_n)_{n=0}^{\infty}\in \F\) is
measured by the Hamiltonian $H_f$ defined by
\begin{equation}\label{hbl:Hf}
\begin{split}
  (H_f\ph)_{0}&=0\\
  (H_f\ph)_{n}(k_1,\dots,k_n) &= \sum_{j=1}^n
  \omega(k_j)\ph_n(k_1,\dots,k_n),\qquad n\geq 1,
\end{split}
\end{equation}
where \(\omega(k)=|k|\). The domain of $H_f$ is the largest set of
vectors for which \eqref{hbl:Hf} defines a vector in $\F$.

The interaction between photons and electrons comes about in a
process of creation and annihilation of photons. To describe it
mathematically, creation and annihilation operators are needed.
Given $h\in L^2(\R^3)$ and $\ph\in \Fin$ we define \(a^{*}(h)\ph\)
by
\begin{equation*}
  [a^{*}(h)\ph]_n = \sqrt{n} {\mathcal S_n}(h\otimes \ph_{n-1}).
\end{equation*}
The operator $a^{*}(h)$ is called a \emph{creation operator}. It
adds a photon with wave function $h$ to the state $\ph$. The
\emph{annihilation operator} $a(h)$ is the adjoint of the closure
of $a^{*}(h)$. These operators satisfy the canonical commutation
relations
\begin{equation}\label{hbl:CCR}
  [a(g),a^{*}(h)] = (g,h),\qquad [a^{\sharp}(g),a^{\sharp}(h)]=0.
\end{equation}
A further important operator on $\F$ is the number operator $N_f$,
defined by
\begin{equation*}
  (N_f\ph)_n = n\ph_n
\end{equation*}
and $D(N_f)=\{\ph\in \F: \sum n^2\|\ph_n\|^2\}<\infty$.

A state of the composed system of electron and photons is
described by a vector $\Psi\in\Hel\otimes\F$, that is, by a
sequence $(\psi_n)_{n=0}^{\infty}$ where $\psi_n$ is a square
integrable function
\begin{equation*}
   \psi_n(x,k_1,\dots, k_n),
\end{equation*}
describing a state of one electron and $n$ photons. It is often
helpful to use that $\Hel\otimes \F\simeq L^2(\R^{3};\F)$ and to
consider $\Psi$ as a square integrable function $x\mapsto \Psi(x)$
with values in $\F$. Then $\|\Psi(x)\|_{\F}^2$ is the probability
density for finding the electron at position $x\in\R^3$.

For the generator of the time-evolution $t\mapsto \Psi_t$ we
choose the Hamiltonian
\begin{equation}\label{hbl:ham}
  H\equiv H_g = \Hamel\otimes 1 + 1\otimes H_f + g\Hint,
\end{equation}
the interaction $\Hint$ being given by
\begin{align*}
  (\Hint\Psi)(x) &= [a(G_x)+a^{*}(G_x)]\Psi(x) \\
  G_x(k) &= e^{-ik\cdot x} \kappa(k),
\end{align*}
where $g\in\R$ and $\kappa\in C_0^{\infty}(\R^3)$. It is easy to
prove that $\Hint$ is operator-bounded with respect to $H_{g=0}$
with bound zero, and hence, for every $g\in\R$, $H_{g}$ is bounded
below and self-adjoint on the domain of $H_{g=0}$ by the
Kato-Rellich theorem~\cite{ReedSimon2}.

Both $g$ and $\kappa$ measure the strength of interaction between
electron and photons. Given $\kappa$, some of the results in the
following sections hold for $|g|$ small enough only, others for
small $|g|\neq 0$. The value $\kappa(k)$ of the \emph{form-factor}
$\kappa$ measures the strength of interaction between electron and
radiation with wave vector $k$. There is no interaction for $k$
outside the support of $\kappa$, which is the case, e.g., for
$|k|$ larger than the \emph{ultraviolet cutoff}
$\Lambda:=\sup\{|k|:\kappa(k)\neq 0\}$.

We reiterate that this model is a \emph{caricature} of the
standard model of quantum electrodynamics for atoms and molecules
interacting with quantized radiation. With QED it has in common
that it describes a non-relativistic particle, the electron,
interacting with massless relativistic bosons in a momentum
conserving process of creation and annihilation of such bosons.

It would make our toy model physically more realistic if we
assumed $\kappa(k)\sim |k|^{-1/2}$ for small $|k|$. But then $H$
has no ground state \cite{Lorinczi2002}, a problem that does not
occur in the standard model of non-relativistic QED. Therefore we
assume that $\kappa$ is non-singular near $k=0$.

\subsection{Spectrum and Eigenfunctions of $\Hamel$}
\label{hbl-sec:schroed}

As a preparation for the following sections we recall a few facts
concerning the spectrum of Schr\"odinger operators
$\Hamel=-\Delta+V$ and the decay of their eigenfunctions.

Suppose that $V$ satisfies assumption \eqref{hbl:V1} and let
$\Hamel$ be defined in terms of the Friedrichs' extension of the
symmetric operator $-\Delta+V$ on $C_0^{\infty}(\R^{3})$. Let
$D_R=C_0^{\infty}(|x|>R)$, the space of smooth, compactly
supported functions with support outside the ball $B_R(0)$, and
let
\begin{equation}\label{hbl:Sigma-el}
  \Sel := \lim_{R\to\infty}\left(\inf_{\ph\in
  D_R,\,\|\ph\|=1}\sprod{\ph}{\Hamel\ph}\right).
\end{equation}
By a theorem due to Arne Persson \cite{Persson, Agmon}
\begin{equation}\label{hbl:persson}
   \Sel = \inf\sigma_{ess}(\Hamel)
\end{equation}
where $\sigma_{ess}(\Hamel)$ denote the essential spectrum of
$\Hamel$, i.e. the complement, within the spectrum, of the
isolated eigenvalues of finite multiplicity. For the Coulomb
potential \eqref{hbl:coulomb}, $V_Z(x)\to0$ as $|x|\to\infty$ and
hence $\Sel=0$. Furthermore $\sigma_{ess}(\Hamel)=[0,\infty)$, and
by a simple variational argument, $\Hamel$ has infinitely many
eigenvalues below $0$ \cite{ReedSimon4}.

Eigenfunctions of $\Hamel$ with energy below $\Sel$ decay
exponentially with increasing $|x|$: for every eigenvalue $E<\Sel$
and every $\beta>0$ with $E+\beta^2<\Sel$ there exists a constant
$C_{\beta}$ such that
\begin{equation}\label{hbl:p-wise-decay}
  |\psi(x)| \leq C_{\beta} e^{-\beta |x|},\qquad \text{a.e. on}\
  \R^3
\end{equation}
for all normalized eigenfunctions that belong to $E$. Of course,
the actual decay of $\psi$ will not be isotropic unless $V$ is
spherically symmetric. There is the better, but non-explicit,
bound $|\psi(x)| \leq C_{\eps}e^{-(1-\eps)\rho(x)}$ where
$\rho(x)$ is the geodesic distance from $x$ to the origin with
respect to a certain metric $ds^2=c_E(x/|x|)dx^2$ in $\R^3$
\cite{Agmon}. Since $c_E(x/|x|)\geq \Sel-E$ (if $\Sel<\infty$),
the isotropic bound \eqref{hbl:p-wise-decay} follows from this
stronger result.

For proving \eqref{hbl:p-wise-decay} it suffices to show that
\begin{equation}\label{hbl:schroed-decay}
  e^{\beta|\cdot|}\psi \in L^2(\R^{3})
\end{equation}
whenever $E+\beta^2<\inf\sigma_{ess}(\Hamel)$. The point-wise
bound \eqref{hbl:p-wise-decay} then follows from a general result
on point-wise bounds for (weak) solutions of second order elliptic
equations \cite{Agmon, GilTru}. The $L^2$-bound
\eqref{hbl:schroed-decay}, in turn, is easily derived from the
characterization \eqref{hbl:Sigma-el}, \eqref{hbl:persson} for
$\inf\sigma_{ess}(\Hamel)$ \cite{HunzikerSigal2000}.

We now turn again to $H_g$, the Hamiltonian~\eqref{hbl:ham}
describing matter and radiation. Most properties of $H_g$ to be
discussed in the following sections hold for all $g\in \R$,
including $g=0$. Hence they generalize properties of
\begin{equation*}
 H_{g=0} = \Hamel\otimes 1 + 1 \otimes H_f.
\end{equation*}
By general spectral theory \(\sigma(H_0)=\overline{\sigma(\Hamel)+
\sigma(H_f)}\). Furthermore, it is easy to see from the definition
of $H_f$ that $\sigma(H_f)=[0,\infty)$ and that $0$ is the only
eigenvalue of $H_f$, the vacuum $\vac$ being its eigenvector. It
follows that $\sigma(H_0)=[\inf\sigma(\Hamel),\infty)$ and that
$H_0$ and $\Hamel$ have the same eigenvalues. The corresponding
eigenvectors are the products $\psi\otimes \vac$, where $\psi$ is
an eigenvector of $\Hamel$.

\subsection{Physical Phenomena and Mathematical Description}
\label{hbl-sec:physics}

The experimental evidence on isolated atoms in contact with
radiation is easiest described in an idealized setup or
``Gedankenexperiment". Consider an atom in a universe that is
otherwise free of matter. For simplicity, we assume that the atom
has only one electron. There may be radiation near the atom
initially but no ``external fields'' or sources of radiation shall
be present. Independent of its initial state, this system of atom
\emph{and} radiation will eventually approach one of only two
qualitatively distinct final states: the ``bound state" or the
``ionized state".

In the bound state the atom is in its ground state, the state of
least energy, where the electron is confined to within a small
neighborhood of the nucleus. All excess energy has been radiated
off. This radiation is very far away from the atom and escaping at
the speed of light.

In the ionized state the electron and the nucleus are spatially
separated with increasing distance. In addition, there may be
radiation going off to infinity.

Of course, the ionized state can only be attained if the total
energy of matter and radiation initially is hight enough to
overcome the attraction between nucleus and electron. High energy
however, does not guarantee ionization, as the excess energy may
just as well turn into radiation. On the other hand, if the total
energy initially is not sufficient for ionization, then the atom
will certainly relax to its ground state. Other conceivable
scenarios, like relaxation to a stationary state with non-minimal
energy, or the permanent radiation without total loss of the
excess energy have never been observed. Bohr's \emph{stationary
states}, with the exception of the ground state, are unstable, and
radiation is only emitted in the very short period of transition
to the ground state.

The goal is to give a proof of the phenomenon of relaxation to the
ground state in the model introduced in the previous section. In
view of the experimental evidence described above, we expect that
this model has the following mathematical properties.

\begin{description}
\item[\em Existence of the ionization threshold.] There exists a
threshold energy $\Sigma\geq \inf\sigma(H)$, such that the
electrons described by states in the spectral subspace
$E_{(-\infty,\Sigma)}(H)\H$ are well localized near the nuclei,
while states with energy above $\Sigma$ may be ionized.
\item[\em Existence of a ground state.] There exists a state of least
energy (ground state), or, equivalently, $\inf\sigma(H)$ is an
eigenvalue of $H$.
\item[\em Absence of excited stationary states.] The operator $H$ has no
eigenvalues above $\inf\sigma(H)$.
\item[\em Asymptotic completeness of Rayleigh scattering (ACR).] In the limit $t\to\infty$
the time evolution $e^{-iHt}\Psi$ of every state $\Psi\in
E_{(-\infty,\Sigma)}(H)\H$ is well approximated by a superposition
of states of the form
\begin{equation}\label{hbl:phys-ACR}
  a^{*}(h_{1,t})\cdot\cdots\cdot a^{*}(h_{n,t}) e^{-iE_{0}t}\Psi_0,
\end{equation}
where \(h_{i,t}(k)=e^{-i|k|t}h_{i}(k)\), $\Psi_0$ is a normalized
ground state of $H$ and $E_0$ is its energy. The vector
\eqref{hbl:phys-ACR} describes a state composed of the atom in its
ground state and $n$ freely propagating photons with wave
functions $h_{1,t},\dots,h_{n,t}$. In the limit $t\to\infty$ they
will be far away from the atom.
\end{description}

The papers \cite{Griesemer2002, GLL, FGS1, FGS2} are devoted to
proving the above four properties for the standard model of QED
was well as for the model introduced in
Section~\ref{hbl-sec:model}. One exception is asymptotic
completeness for Rayleigh scattering where the results concern the
model of Section~\ref{hbl-sec:model} only, with the additional
important simplification that $\kappa(k)=0$ for $|k|\leq \sigma$
where $\sigma>0$ is arbitrarily small but positive. This
assumptions that photons with energy near zero don't interact with
the electron is usually referred to as an \emph{infrared cutoff},
IR-cutoff, for short. Sometimes the constant $\sigma$ is called
infrared cutoff as well. The significance of the IR-cutoff is that
it allows us to control the number of bosons that are being
produced in the course of time: the number of bosons with energy
below $\sigma$ stays constant and the number of bosons with energy
above $\sigma$ can be bounded from above in terms of the total
energy. The assumption of an IR-cutoff is \emph{not} expected to
be necessary for the validity of asymptotic completeness of
Rayleigh scattering as formulated above; but as of now, no
convincing mathematical argument is known that would substantiate
this believe.

The above list of physical phenomena is limited to the coarsest
properties of atoms interacting with radiation. Even below the
ionization threshold, there are many other phenomena that are
worth rediscovering in the standard model, and in part this has
already been done. Most important, perhaps, are the occurrence of
sharp lines in the spectrum of the emitted radiation (Bohr
frequencies), the resonances in lieu of Bohr's stationary state
and their extended life time, and the correspondence principle at
energies near the ionization threshold. We shall come back to some
of these phenomena in later sections, when we review known results
or comment on open problems. One must keep in mind, however, the
limitations of our model. Quantitative predictions will be of
limited accuracy when relativity or high-energy photons play a
significant role.

\subsection{Exponential Decay and Ionization Threshold}
\label{hbl-sec:expo}

The \emph{ionization threshold} $\Sigma$ of an atom or molecule
with only one electron is the least energy that this system can
achieve in a state where the electron has been moved ``infinitely
far away'' from the nuclei. The electron is outside the ball $|x|<
R$ with probability one, if its wave function $\Psi(x)$ vanishes
in this ball. Therefore we define
\begin{equation}\label{hbl:threshold}
\Sigma = \lim_{R\to\infty}\left(\inf_{\Psi\in
 D_R,\|\Psi\|=1}\sprod{\Psi}{H\Psi}\right)
\end{equation}
where
\begin{equation*}
  D_R = \{\Psi\in D(H)|\Psi(x)=0\ \text{if}\ |x|< R\}.
\end{equation*}
Note the analogy with Persson's characterization
\eqref{hbl:Sigma-el} of $\inf\sigma_{ess}(\Hamel)$. Here, however,
$\Sigma\neq\inf\sigma_{ess}(H)$ unless $\Sigma=\inf\sigma(H)$. In
general $\Sigma\geq \inf\sigma(H)$ and, if $V(x)\to 0$ as
$|x|\to\infty$, then \(\Sigma=\inf\sigma(H-V)\), which is greater
than $\inf\sigma(H)$ for $V=V_Z$ \cite{GLL} (see also
Section~\ref{hbl-sec:gs}).

According to \eqref{hbl:threshold}, the electron described by a
state $\Psi$ with energy below $\Sigma$ cannot be arbitrarily far
away from the origin. In fact, by \cite{Griesemer2002}, for all
$\lambda,\beta\in \R$ with $\lambda+\beta^2<\Sigma$
\begin{equation}\label{hbl:decay}
  \big\| (e^{\beta|x|}\otimes 1) E_{\lambda}(H)\big\| <\infty.
\end{equation}
This shows that the probability(-density) $\|\Psi(x)\|^2_{\F}$ to
find the electron at the point $x$ decays exponentially fast, as
$|x|\to\infty$, at least in the averaged sense
\begin{equation}\label{hbl:L2-decay}
  \int e^{2\beta|x|} \|\Psi(x)\|^2_{\F}\, dx <\infty.
\end{equation}

There is an obvious similarity between \eqref{hbl:L2-decay} and
\eqref{hbl:schroed-decay} that is not accidental. In
\cite{Griesemer2002} the bound \eqref{hbl:decay} is derived from
an abstract result for semi-bounded self-adjoint operators $H$ in
Hilbert spaces of the form $L^2(\R^n)\otimes\F$, where $\F$ is an
arbitrary complex Hilbert space. The only assumptions are that
\(fD(|H|^{1/2})\subset D(|H|^{1/2})\) and that
\begin{equation}\label{hbl:double-com}
  f^2 H + H f^2 -2fHf = -2|\nabla f|^2
\end{equation}
for smooth, bounded functions $f(x)$ with bounded first
derivatives. Equation~\eqref{hbl:double-com} holds for
$H=-\Delta\otimes 1$ and since the left-hand side of
\eqref{hbl:double-com} formally equals $[f,[f,H]]$, it follows
that \eqref{hbl:double-com} holds for are large class of
self-adjoint operators $H$ whose principal symbol is given by the
Laplacian. The result \eqref{hbl:schroed-decay} thus emerges as a
special case of \eqref{hbl:decay}.

Even though the above assumptions on $H$ are largely independent
of $H_f$ and $g\Hint$, the result depends on these operators! The
binding energy $\Sigma-\inf\sigma(H)$ depends of $g$ and hence so
does the decay rate one obtains for the ground state.

It is well known, since the work of Agmon \cite{Agmon}, that
eigenfunctions of second order elliptic equations decay
exponentially in energetically forbidden regions, that is, in
regions where the differential operator, as a quadratic form, is
strictly larger than the eigenvalue \cite{Agmon}. The result
\eqref{hbl:decay} shows that this idea can be brought to bear in a
much more general framework, including many models of
non-relativistic QED. It is clear that our proof of
\eqref{hbl:decay} can be generalized, along the lines of
\cite{Agmon}, to yield non-isotropic bounds, as well as
exponential bounds for QED-models where $\Hamel$ is a more
general, uniformly elliptic second order differential operator.

We owe the strategy for proving \eqref{hbl:decay} to Bach et al.
\cite{BFS1}, where this bound is established for $|g|$
sufficiently small and \(\lambda+\beta^2<\Sel-\const |g|\).

\subsection{Existence of a Ground State}
\label{hbl-sec:gs}

By the main result of \cite{GLL}, $\inf\sigma(H)$ is an eigenvalue
of $H$ whenever
\begin{equation}\label{hbl:bind}
  \inf\sigma(H) < \Sigma.
\end{equation}
This reduces a difficult spectral problem to a variational
problem: the problem of finding a state $\Psi\in D(H)$ with
\(\sprod{\Psi}{H\Psi}<\Sigma\sprod{\Psi}{\Psi}\). There are two
important classes of potentials $V$ for which this variational
problem can be solved without much effort: if $V(x)\to\infty$ as
$|x|\to\infty$, then obviously $\Sigma=\infty$ and hence
$\inf\sigma(H)<\Sigma$. On the other hand, if $V(x)\to 0$ as
$|x|\to\infty$ then $\Sigma=\inf\sigma(H-V)$ where the nuclei are
removed in $H-V$. Using the translation invariance of $H-V$ one
shows that
\begin{equation*}
\inf\sigma(H)\leq \inf\sigma(H-V) + \inf\sigma(-\Delta+V).
\end{equation*}
It follows that \(\inf\sigma(H)<\Sigma\) whenever $V(x)\to 0,\
(|x|\to\infty)$ and $\inf\sigma(-\Delta+V)<0$. Since this is the
case for the Coulomb potential $V=V_Z$, all one-electron atoms and
molecules have a ground state.

We next sketch the proof that \eqref{hbl:bind} guarantees
existence of a ground state. To begin with we recall that a
hypothetical eigenvector $\Psi$ of $H$, with eigenvalue
$\inf\sigma(H)$, minimizes the quadratic form $\Psi\mapsto
\sprod{\Psi}{H\Psi}$ subject to the constraint $\|\Psi\|=1$. It is
thus natural to establish existence of $\Psi$ by proving relative
compactness for a suitable minimizing sequence. The problem with
this approach is that a generic minimizing sequence will tend
weakly to zero. The fact that $\inf\sigma(H)$ belongs to the
essential spectrum alone implies that there are infinitely many
energy minimizing sequences with this defect. Our task is thus,
first, to choose a suitable minimizing sequence, and second, to
prove its relative compactness.

We choose the elements of our minimizing sequence to be the ground
states $\Psi_m$ of modified Hamiltonians $H_m$ ($m\to 0$) in which
the photon energy $\omega(k)$ is altered to be
$\omega_m(k)=\sqrt{k^2+m^2}$. That is, we give the photons a
positive mass $m$. For $m$ small enough the binding assumption
\eqref{hbl:bind} is inherited by $H_m$, which we use to show that
$\inf\sigma(H_m)$ is indeed an eigenvalue. As a matter of fact,
$\inf\sigma(H_m)$ is separated from the essential spectrum of
$H_{m}$ by a gap of size $m$ \cite{FGS2}. The sequence of ground
states $(\Psi_m)_{m>0}$ is a minimizing sequence for $H$ that can
be assumed to be weakly convergent. It remains to show that the
weak limit is not the zero vector in $\H$.

We first argue that it suffices to prove relative compactness of
the sequence of $L^2$-functions $\psi_{m,n}(x,k_1,\dots,k_n)$
restricted to large balls $B\subset \R^{(3+3n)}$. This follows
from the exponential decay w.r.to $x$, from
$\psi_{m,n}(x,k_1,\dots,k_n)=0$ if $|k_i|>\Lambda$, and from the
bound \(\sup_m\sprod{\Psi_m}{N_f\Psi_m}<\infty\). Then we use the
compactness of the embedding
\begin{equation*}
  W^{1,p}(B) \hookrightarrow L^2(B),\qquad \text{for}\
  2>p>\frac{2\cdot (3+3n)}{2+(3+3n)}
\end{equation*}
due to Rellich-Kondrachov \cite{Adams}. We thus need to show that
\(\sup_{m}\|\nabla\psi_{m,n}\|_p<\infty\), which we derive from a
refinement of the argument that we used to prove the bound on
\(\sprod{\Psi_m}{N_f\Psi_m}\), and from the $H$-boundedness of
$-\Delta_x$.

There was a large number of previous papers on the existence of a
ground state for models similar to the one discussed here
\cite{AraiHirokawa1997, AraiHirokawa2000, BFS1, BFS2, Gerard2000,
Hiroshima1999, Hiroshima2000a, Spohn1998}. Among these, the best
result is due to Bach et al \cite{BFS2}. It established existence
of a ground state when $\inf\sigma(\HamNel{N})$ is an isolated
eigenvalue and the fine-structure constant $\alpha$ is small
enough. Most importantly, in this paper for the first time
existence of a ground state is proven in the standard model
without an IR-regularization.

\subsection{Relaxation to the Ground State is a Scattering
Phenomenon} \label{hbl-sec:relax}

As discussed in Section~\ref{hbl-sec:physics} every state $\Psi\in
\ran E_{(-\infty,\Sigma)}(H)$ is expected to ``relax to the ground
state by emission of photons''. In mathematical terms, this means
that $e^{-iHt}\Psi$, in the distant future, $t\to \infty$, is well
approximated in norm by a linear combination of vectors of the
form
\begin{equation}\label{hbl:relax}
 a^{*}(h_{1,t})\dots a^{*}(h_{n,t}) e^{-iE_0 t}\Psi_0
\end{equation}
where $h_{i,t}(k)= e^{-i\omega t}h_i(k)$, $\Psi_0$ is the ground
state, and $E_0$ is its energy. This has a chance to be correct
only if $H$ has no other eigenvalues below $\Sigma$. If it has,
then \emph{relaxation to a bound state} may occur, which means
that $\Psi_0$ in \eqref{hbl:relax} may be \emph{any} eigenvector
of $H$ with eigenvalue below $\Sigma$. In this weaker form, the
above assertion is called \emph{Asymptotic Completeness for
Rayleigh scattering}. The problem of proving absence of excited
eigenvalues is independent of the scattering problem and its
discussion is deferred to a later section.

Before proving completeness of the scattering states one needs to
address the problem of their existence. An example of a
\emph{scattering state} is a vector $\Psi_{+}\in\H$ for which
there exist $n\geq 1$ photons $h_1,\dots, h_n$ and an eigenvector
$\Psi$, $H\Psi=E\Psi$ such that
\begin{equation*}
  e^{-iHt}\Psi_{+} \simeq a^{*}(h_{1,t})\dots a^{*}(h_{n,t}) e^{-iE
  t}\Psi,\qquad t\to\infty
\end{equation*}
in the sense that norm of the difference vanishes in the limit
$t\to\infty$. The scattering state $\Psi_{+}$ is said to
\emph{exists} if the limit
\begin{equation}\label{hbl:scat-exist}
  \Psi_{+} =\lim_{t\to\infty} e^{iHt} a^{*}(h_{1,t})\dots a^{*}(h_{n,t})
  e^{-iEt}\Psi\qquad t\to\infty
\end{equation}
exists. Let $\H_{+}$ denote the closure of the space spanned by
vectors $\Psi_{+}$ of the form \eqref{hbl:scat-exist}. All
elements of $\H_{+}$ are called scattering states and asymptotic
completeness of Rayleigh scattering is the property that
\begin{equation}\label{hbl:ac2}
  \H_{+}\supset \ran E_{(-\infty,\Sigma)}(H).
\end{equation}
Existence of scattering states is established in \cite{FGS1}, and
\eqref{hbl:ac2} is proven in \cite{FGS2} for the model introduced
in Section~\ref{hbl-sec:model}, assuming either an infrared cutoff
on the interaction or that the photon dispersions relation
$\omega(k)$ is bounded from below by a positive constant, which
excludes $\omega(k)=|k|$. The latter assumption serves the same
purpose as the infrared cutoff, and it is satisfied, e.g., for
massive bosons where \(\omega(k)=\sqrt{k^2+m^2}\) with $m>0$. As
of today, there is no proof of \eqref{hbl:ac2} without a form of
infrared cutoff or another drastically simplifying assumption
\cite{Arai83,Spohn}.

\subsection{Existence of Scattering States}
\label{hbl-sec:scatex}

Generalizing \eqref{hbl:scat-exist}, we ask whether the limits
\begin{equation}\label{hbl:n-phot}
  \Psi_{+} = \lim_{t\to\infty} e^{iHt}a^{\#}(h_{1,t})\dots a^{\#}(h_{n,t}) e^{-iH t}\Psi
\end{equation}
exist for given $\Psi\in \H$, and \(h_i\in L^2(\R^3)\), where
$a^{\#}(h_{i,t})$ is a creation or an annihilation operator. The
scattering states $\Psi_{-}$ obtained in the limit $t\to-\infty$
are physically interesting as well, but the problem of their
existence is mathematically equivalent to the existence of
\eqref{hbl:n-phot}. Beginning with the easiest case, $n=1$, let us
ask whether the limits
\begin{equation}\label{hbl:asy-op}
  a_{+}^{\#}(h)\Psi = \lim_{t\to\infty}
  e^{iHt}a^{\#}(h_t)e^{-iHt}\Psi
\end{equation}
exist. If the photons are \emph{massless}, as they are in nature,
the answer depends on the electron dispersion relation and on the
energy distribution of $\Psi$. For \emph{massive} photons,
however, $a_{+}^{\#}(h)\Psi$ exists for all $\Psi\in D(H)$ and all
\(h\in C_0^{\infty}(\R^3)\), and the proof is short and easy
\cite{H-K}: by the Cauchy criterion the limit \eqref{hbl:asy-op}
exists if the time derivative of the right-hand side is
(absolutely) integrable. A straightforward computation using
$a^*(h_t)=e^{-iH_ft}a^*(h)e^{iH_ft}$ and \eqref{hbl:CCR} shows
that
\begin{equation}\label{abs-int}
 \int_1^{\infty} \left\| \frac{d}{dt} e^{iHt}a^{*}(h_t)e^{-iHt}\Psi \right\| dt
 = \int_1^{\infty} \left\| (G_x,h_t)e^{-iHt}\Psi \right\| dt
\end{equation}
where
\begin{equation}\label{phase-int}
  (G_x,h_t) = \int e^{ik\cdot x-i\omega t}\overline{\kappa(k)}h(k)\,
  dk
\end{equation}
and \(\omega(k)=\sqrt{k^2+m^2}\). Since the Hessian of $\omega$ is
strictly positive,
$$
  \sup_x|(G_x,h_t)| \leq \const\ t^{-3/2}, \qquad t\geq 1
$$
by a standard result on oscillatory integrals \cite{ReedSimon3}.
Hence \(\|(G_x,h_t)e^{-iHt}\Psi\|\leq\const\ t^{-3/2}\) and
\eqref{abs-int} is finite which proves the existence of
$a_{+}^{*}(h)\Psi$. For \emph{massless} photons, however,
$$
  \sup_x|(G_x,h_t)| \sim \const\ t^{-1}, \qquad t\geq 1
$$
which is not integrable and we need to estimate the integrand of
\eqref{abs-int} more carefully. To begin with, we note that the
phase in \eqref{phase-int} is non-stationary away from the ``wave
front'' $|x|=t$. Hence
\begin{equation*}
  \sup_{x:||x|/t-1|\geq \eps}|(G_x,h_t)| \leq \frac{C_n}{t^n}, \qquad t\geq 1
\end{equation*}
for every integer $n$ \cite{ReedSimon3}, and it remains to
estimate
\begin{equation}\label{hbl:el-prop-est}
  \int_0^{\infty} \frac{dt}{t} \left\|\chi_{[1-\eps,1+\eps]}(|x|/t)
  e^{-iHt}\Psi\right\|.
\end{equation}
Finiteness of \eqref{hbl:el-prop-est} requires, in particular,
that the electrons do not propagate at the speed of light, which
is true in nature, but not precluded for the dynamics generated by
the non-relativistic Schr\"odinger operator $\Hamel$. The easiest
case occurs when the electrons are in a bound state $\Psi\in \ran
E_{\lambda}(H)$, $\lambda<\Sigma$, where, by \eqref{hbl:decay},
\begin{equation}\label{hbl:t-uni-decay}
   \sup_{t}\|e^{\beta|x|}e^{-iHt}\Psi\| \leq
   \left\| e^{\beta|x|}E_{\lambda}(H) \right\| < \infty
\end{equation}
for some $\beta>0$. Then obviously the integrand in
\eqref{hbl:el-prop-est} decays exponentially in time, and hence
the limit \eqref{hbl:asy-op} exists. If $\Psi$ is not in a bound
state but its energy is insufficient for an electron to reach the
speed $1-\eps$, then \eqref{hbl:el-prop-est} is still finite, at
least if $H$ is the Hamiltonian \eqref{eq:qed1} of the standard
model \cite{FGS1}. More precisely, \eqref{hbl:el-prop-est} is
finite for all $\Psi$ in a dense subspace of $\ran E_{\lambda}(H)$
with $\lambda <\Sigma +m/2$, and for $\eps$ in
\eqref{hbl:el-prop-est} small enough. Here $m/2=mc^2/2$ is the
non-relativistic kinetic energy of a particle at the speed of
light. The assumption $\lambda <\Sigma +m/2$ thus ensures that no
electron can reach the speed of light.

The asymptotic field operators have the important property that
\begin{equation}\label{hbl:add-energy}
\begin{split}
  a_{+}^{*}(g)\ran E_{\lambda}(H) &\subset \ran E_{\lambda+M}(H)\\
  a_{+}(h)\ran E_{\lambda}(H) &\subset \ran E_{\lambda-m}(H)
\end{split}
\end{equation}
if \(\supp(g)\subset \{k: |k|\leq M\}\) and \(\supp(h)\subset \{k:
|k|\geq m\}\). Using \eqref{hbl:add-energy} and the existence of
the limit \eqref{hbl:asy-op} we prove existence of the limit
\eqref{hbl:n-phot} and that
\begin{equation}\label{hbl:n-asy-op}
  \lim_{t\to\infty} e^{iHt}a^{\#}(h_{1,t})\dots a^{\#}(h_{n,t}) e^{-iH t}\Psi
  =a_{+}^{\#}(h_1)\dots a_{+}^{\#}(h_n)\Psi
\end{equation}
if \(\psi\in \ran E_{\lambda}(H)\), \(\lambda+\sum_j M_j <
\Sigma\), where \(M_j=\sup\{|k|:h_j(k)\neq 0\}\) and the sum
$\sum_j M_j$ extends over all creation operators in
\eqref{hbl:n-asy-op}. The main technical difficulty in this last
step is the unboundedness of the asymptotic field operators
\eqref{hbl:asy-op}.

From the above discussion it is clear that it is physically more
sensible to describe the time evolution of the electron by a
relativistic Hamiltonian such as
\begin{equation}\label{hbl:rel-at}
  \Hamel^{\rm rel} = \sqrt{-\Delta +1} + V(X),
\end{equation}
in place of \eqref{hbl:schroed}. Then \eqref{hbl:el-prop-est} is
finite for all $\Psi$ in any spectral subspace
\(E_{\lambda}(H)\H\) with  \(\lambda<\infty\). Hence the
asymptotic operators $a_{+}^{\#}(h)$ exist on a dense subspace of
$\H$. The main results in \cite{FGS1, FGS2, FGS3} apply to both
electron-Hamiltonian, \eqref{hbl:schroed} and \eqref{hbl:rel-at}.

To conclude this discussion of scattering states we remark that
for Rayleigh scattering it suffices to prove existence of the
asymptotic field-operators $a_{+}^{\#}(h)$ on $\ran
E_{(-\infty,\Sigma)}(H)$, which follows from
\eqref{hbl:t-uni-decay}. The improved results discussed thereafter
are important in the study of photon scattering at a free electron
(Compton scattering) \cite{FGS3}.

\subsection{Asymptotic Completeness}
\label{hbl-sec:ac}

A characterization of ACR that is mathematically more convenient
than \eqref{hbl:ac2} is achieved by mapping the freely propagating
photons $h_{i,t}$ in \eqref{hbl:scat-exist} into an auxiliary Fock
space $\F$. We attach $\F$ to the Hilbert space $\H$ by defining
an extended Hilbert space \(\HxF=\H\otimes \F\). The appropriate
time evolution on $\HxF$ is generated by the extended Hamiltonian
\(\Hex = H\otimes 1 + 1\otimes H_f \). Furthermore we define an
identification operator \(I: D\subset \HxF \to \H\) on a dense
subspace $D$ of $\HxF$ by
\begin{align*}
I\, \Psi\otimes \vac &= \Psi\\
I\, \Psi \otimes a^{*}(h_1)\cdots a^{*}(h_n)\vac
&=a^{*}(h_1)\cdots a^{*}(h_n)\Psi
\end{align*}
and linear extension. Since \(e^{-iH_f t} \vac = \vac\) and
\(a^{*}(h_t)=e^{-iH_f t} a^{*}(h) e^{iH_f t}\) we can use $I$ to
write
$$
  a^{*}(h_{1,t})\cdots a^{*}(h_{n,t})e^{-iHt}\Psi=
  I e^{-i\Hex t}\big[\Psi\otimes a^{*}(h_1)\cdots
  a^{*}(h_n)\vac\big].
$$
If $\Psi$ is an eigenvector of $H$ and $P_{B}$ denotes the
orthogonal projection onto the closure of the span of all
eigenvectors, it follows that
\begin{align*}
  a_{+}^{*}(h_1)\cdots a_{+}^{*}(h_n)\Psi &=
  \lim_{t\to\infty}e^{iHt} a^{*}(h_{1,t})\cdots
  a^{*}(h_{n,t})e^{-iHt}\Psi\\
  &= \Omega_{+}\big[\Psi\otimes a^{*}(h_1)\cdots a^{*}(h_n)\vac\big]
\end{align*}
where
$$
  \Omega_{+} = s-\lim_{t\to\infty} e^{iHt} I e^{-i\Hex t}
  P_{B}\otimes 1
$$
is the \emph{wave operator}. Thus existence of scattering states
becomes equivalent to existence of the wave operator $\Omega_{+}$,
and, since $\ran \Omega_{+}=\H_{+}$, asymptotic completeness as
defined in \eqref{hbl:ac2} becomes
\begin{equation}\label{hbl:ACRwave}
  E_{(-\infty,\Sigma)}(H)\subset \ran \Omega_{+}.
\end{equation}
(It turns out that $\Omega_{+}$ is a partial isometry and hence
$\ran \Omega_{+}$ is closed.)

The reader familiar with quantum mechanical scattering theory is
cautioned not to think of ACR as a form of asymptotic completeness
for potential scattering. The comparison dynamics generated by
$\Hex$ is \emph{not} the free dynamics for all bosons. The bosons
in the first factor of $\HxF$ still fully interact with the
electrons. Rayleigh scattering is more similar to $N$-body quantum
scattering with the additional complication that the number of
particles is fluctuating.

Asymptotic completeness of Rayleigh scattering, as described in
Section~\ref{hbl-sec:relax}, makes two assertions. First, any
initial state $\Psi\in\ran E_{(-\infty,\Sigma)}(H)$, that is not
an eigenvector of $H$, in the course of time will relax to a bound
state by emission of photons. Second, the asymptotic dynamics of
the emitted radiation is well approximated by the free photon
dynamics. Our proof of ACR contains two main technical ingredients
that address these issues. In both of them we need to assume that
either the photons are massive, i.e., $\omega(k)=\sqrt{k^2+m^2}$,
or that an infrared cutoff $\sigma>0$ is imposed on the
interaction. In the second case $m=\sigma/2$ in the following.

Our first main ingredient is an estimate on the ballistic spacial
expansion of the system for states with energy distribution away
from $S=\sigma_{pp}(H)+\N m$. We show that $S$ is closed and
countable and that for each \(\lambda\in \R\backslash S\) there is
an open interval $\Delta\ni\lambda$ and a positive constant
$C_{\lambda}$ such that
\begin{equation}\label{hbl:ballistic}
  \sprod{\Psi_t}{\dGamma(y^2)\Psi_t} \geq C_{\lambda} t^2,\qquad
  t\to \infty
\end{equation}
for all $\Psi\subset \ran E_{\Delta}(H)$. The proof is based on
the positivity of the commutator obtained by differentiating the
left hand side twice with respect to time. This positive
commutator estimate, often called Mourre estimate, is proven by
induction in energy steps of size $m$ along a strategy very
similar to the proof of the Mourre estimate for $N$-body
Schr\"odinger operators \cite{HunzikerSigal2000}. We generalize
the Mourre estimate in \cite{DG1} to accommodate our model.

The second main ingredient is a \emph{propagation estimate} for
the asymptotic dynamics of escaping photons. Explicitly we show
that
\begin{equation}\label{hbl:propes}
 \int_1^{\infty}\frac{dt}{t} \sprod{\Psi_t}{fF\dGamma(P_t)
 Ff\Psi_t}\leq C\|\Psi\|^2
\end{equation}
for all $\Psi\in\H$, where
$$
  P_t = (\nabla\omega-y/t)\cdot \chi(|y|\geq t^{\delta})(\nabla\omega-y/t)
$$
and $0<\delta <1$. Here $f$ is an energy cutoff,
$F=F(\dGamma(y^2/t^2\lambda^2))$ a space cutoff, and $\lambda>0$ a
parameter that is chosen sufficiently large eventually. The
left-hand side of \eqref{hbl:propes} compares the average photon
velocity, $y/t$, with the group velocity, $\nabla\omega$, for
photons in the region $\{|y|\geq t^{\delta}\}$. This includes all
photons that escape the electron ballistically. The finiteness of
$C$ thus confirms that the dynamics of outgoing radiation is
approaching the free photon dynamics in the limit $t\to \infty$.

Asymptotic completeness had previously been established for a
model with $\Sigma=\infty$ (confined electrons), and massive
photons $\omega(k)=\sqrt{k^2+m^2}$, $m>0$, by Derezi\'nski and
G\'erard \cite{DG1}. The methods in \cite{DG1} could probably be
extended to prove ACR for our system. Instead of doing so, we
chose to give an entirely new prove of AC based on the relatively
elementary propagation estimate \eqref{hbl:propes}, and using
\eqref{hbl:ballistic} as the only dynamical consequence of the
Mourre estimate. Our work is inspired by the Graf-Schenker proof
of asymptotic completeness for $N$-body quantum systems
\cite{GrafSchenker1997}.

\subsection{Absence of Excited States}
\label{hbl-sec:no-ev}

At present the knowledge on absence of eigenvalues above
$\inf\sigma(H_g)$ is far less complete then, e.g., our knowledge
regarding existence of a ground state. Known results on absence of
eigenvalues are derived under the assumption that $g>0$ is small
enough \cite{BFS1, BFS3, BFSS}, and to ensure that no new
eigenvalues emerge near $\inf\sigma(H)$ an infrared cutoff is
imposed \cite{FGS2}. There is a further assumption, the Fermi
golden rule condition, which ensures that eigenvalues of $H_{g=0}$
dissolve for $g\neq 0$. This assumption can be checked in any
explicitly given model.

For the model introduced in Section~\ref{hbl-sec:model}, with the
assumption of an infrared cutoff, the following results hold true.
For any given $\eps>0$ and for $|g|>0$ small enough, depending on
$\eps$,
$$
   \sigma_{pp}(H_g)\cap (\inf\sigma(H_g),\Sel-\eps)=\emptyset,
$$
where $\Sel=\inf\sigma(\Hamel)$ \cite{BFSS, FGS2}. This result,
combined with ACR from the previous section implies that
\(\Ran\Omega_{+}\supset\Ran E_{(\inf\sigma(H), \Sel-\eps)}(H)\)
where the projector $P_B$ in the definition of $\Omega_{+}$ is the
projector onto the ground state. That is, every $\psi\in \Ran
E_{(\inf\sigma(H), \Sel-\eps)}(H)$ relaxes to the ground state in
the sense of Section~\ref{hbl-sec:relax},~\eqref{hbl:relax}.

It has been asserted in \cite{BFS3} that the methods of
\cite{BFS1, BFSS} can be used to show absolute continuity of the
spectrum of $H$ above and away from $\Sel$ for small $|g|$. This
is presumably correct but a proof is missing.

The strategy for proving absence of eigenvalues in a given
spectral interval $\Delta\subset \R$ is clear and simple: One
tries to find a symmetric operator $A$ on $\H$, such that
$$
   E_{\Delta}(H) [iH,A]E_{\Delta}(H) \geq C E_{\Delta}(H)
$$
with a positive constant $C$. Since, formally,
$\sprod{\Psi}{[iH,A]\Psi}=0$ for every eigenvector $\Psi$ of $H$,
it immediately follows that $\sigma_{pp}(H)\cap\Delta=\emptyset$.
The main problems, of course, are to find a suitable conjugate
operator $A$, and to make these formal arguments rigorous.

\subsection{Relaxation to the Ground State}
\label{hbl-sec:relax2gs}

An important consequence of AC for Rayleigh scattering and the
absence of eigenvectors besides a unique ground state $\Psi_0$, is
\emph{relaxation to the ground state}. To explain this let
$\mathcal{A}$ denote the $C^*$-algebra generated by all operators
of the form
$$
   B\otimes e^{i\phi(h)},\qquad B\in \L(\Hel),\ h\in
   C_{0}^{\infty}(\R^3),
$$
where $\phi(h)=a(h)+a^{*}(h)$. We say that $\Psi_t=e^{-iHt}\Psi$
\emph{relaxes to the ground state} $\Psi_0$, if
\begin{equation}\label{hbl:relax-to-gs}
   \lim_{t\to\infty}\sprod{\Psi_t}{A\Psi_t} =
   \sprod{\Psi_0}{A\Psi_0}\sprod{\Psi}{\Psi}
\end{equation}
for all $A\in\mathcal{A}$. Suppose $H$ has a unique ground state
$\Psi_0$ and let $\H_{+}$ denote the space of scattering states
over $\Psi_0$. That is, $\H_{+}$ is the closure of the span of all
vectors of the form
$$
   a^{*}_{+}(h_{1})\cdots a^{*}_{+}(h_{n})\Psi_0.
$$
Then, by a short computation, all states in $\H_{+}$ relax to the
ground state $\Psi_0$ \cite{FGS1}. Since the assumptions of
Section~\ref{hbl-sec:no-ev} imply AC in the form
$$
   \H_{+} \supset \ran E_{(\inf\sigma(H),\Sel-\eps)}(H_g)
$$
for given $\eps>0$ and small enough coupling $|g|$, it follows
that all state in $\ran E_{(\inf\sigma(H),\Sel-\eps)}(H_g)$ relax
to the ground state in the sense of
Equation~\ref{hbl:relax-to-gs}.

\section{N-Electron Atoms and Molecules}
\label{hbl-sec:Nel}

We now briefly describe how the results of the previous sections
are generalized to the case of $N>1$ electrons. For simplicity we
neglect spin and Pauli principle. A (pure) state of $N$ electrons
is described by a vector $\psi\in L^2(\R^{3N})$, and the
Schr\"odinger operator for $N$ electrons in the field of $K$
static nuclei is given by
$$
   \HamNel{N}= \sum_{j=1}^N \left(-\Delta_{x_j}+ V_Z(x_j)\right) +
  \sum_{i<j}\frac{1}{|x_i-x_j|}
$$
where $x_j\in \R^3$ is the position of the $j$th electron. The
coupling of the electrons to the radiation field is done by a
straightforward generalization of \eqref{hbl:ham}. The Hamiltonian
of the entire system is given by
\begin{align*}
H_{g,N} &= \HamNel{N}\otimes 1 + 1\otimes H_f + g\Hint\\
\Hint &= \sum_{j=1}^N \big[a(G_{x_j})+a^{*}(G_{x_j})\big],\qquad
G_{x_j}(k) = e^{-ik\cdot x_j}\kappa(k)
\end{align*}
and acts on $L^2(\R^{3N})\otimes \F$. Again, $H_g$ is self-adjoint
on $D(H_{g=0})$.

\subsubsection{Spectrum and eigenfunctions of $\HamNel{N}$}

Like $\Hamel$, $\HamNel{N}$ is a Schr\"odinger operator of the
general form $-\Delta+V$, and hence $\inf\sigma_{ess}(\HamNel{N})$
is given by Persson's theorem:
\begin{equation*}
  \inf\sigma_{ess}(\HamNel{N})=
  \lim_{R\to\infty}\left(\inf_{\ph\in D_R,\,\|\ph\|=1}\sprod{\ph}{\HamNel{N}\ph}\right)
  =: \SelN{N}
\end{equation*}
where $D_R=C_0^{\infty}(|X|>R)$. Using the decay of the two-body
potentials and the electron-electron repulsion one shows that
$\SelN{N}=\inf\sigma(\HamNel{N-1})$, which leads to
\begin{equation}\label{hbl:HVZ}
  \inf\sigma_{\rm ess}(\HamNel{N}) = \inf\sigma(\HamNel{N-1}),
\end{equation}
a special case of the more general HVZ-Theorem \cite{ReedSimon4,
HunzikerSigal2000}. For $Z>N-1$ the system described by
$\HamNel{N-1}$ has a net positive charge and can bind at least one
more electron. It follows, by a simple variational argument, that
\(\inf\sigma(\HamNel{N})<\inf\sigma(\HamNel{N-1})\), which, by
\eqref{hbl:HVZ} implies that $\inf\sigma(\HamNel{N})$ is an
eigenvalue of $\HamNel{N}$. In fact, $\HamNel{N}$ has infinitely
many (discrete) eigenvalues below \(\inf\sigma_{\rm
ess}(\HamNel{N})\) \cite{ReedSimon4}. The continuous part of the
spectrum of $\HamNel{N}$ is the interval \([\inf\sigma_{\rm
ess}(\HamNel{N}),\infty)\), and this interval may contain further
eigenvalues below $0$ \cite{Thirring3}. For a discussion of the
structure of the continuous spectrum, the reader is referred to
\cite{HunzikerSigal2000}.

The results \eqref{hbl:p-wise-decay}, \eqref{hbl:schroed-decay} on
the decay of eigenfunctions hold for Schr\"odinger operators in
arbitrary dimensions, hence in particular for $\HamNel{N}$.
Better, non-isotropic exponential bounds are known too, but they
are expressed in terms of a geodesic distance $\rho(X)$ w.r.to a
metric in $\R^{3N}$ that depends on the spectra of the
Hamiltonians $\HamNel{k}$, for $k<N$ \cite{Agmon,
HunzikerSigal2000}. Explicit expressions for $\rho$ are known for
$N\leq 3$, and for atoms under the (unproven) assumption that the
ionization energy increases monotonically as the electrons, one by
one, are removed from the atom \cite{CarmonaSimon}.

This concludes our discussion of $\HamNel{N}$ and we return to the
composed system of $N$ electrons and radiation.

\subsubsection{Exponential decay and ionization thresholds}

The ionization threshold $\Sigma$ is the least energy that an atom
or molecule can achieve in a state where one or more electrons
have been moved ``infinitely far away'' from the nuclei. In an
$N$-particle configuration $X\in\R^{3N}$, one or more electrons
are far away from the (static) nuclei if and only if $|X|$ is
large. In this respect there is no difference between $N=1$ and
$N>1$ besides the dimension of the configuration space. Since this
dimension is irrelevant for the proof of \eqref{hbl:decay}, our
result on exponential decay for $N>1$ and its proof are
straightforward generalizations of result and proof for $N=1$. Let
$D_R:= \{\Psi\in D(H)|\, \Psi(X)=0\ \text{if}\ |X|<R\}$ and let
\begin{equation}\label{hbl:threshold-N}
\Sigma_N = \lim_{R\to\infty}\left(\inf_{\Psi\in
 D_R,\|\psi\|=1}\sprod{\Psi}{H_{N}\Psi}\right).
\end{equation}
Then for all real numbers $\lambda$ and $\beta$ with
$\lambda+\beta^2<\Sigma$,
\begin{equation}\label{hbl:decay-N}
  \big\| (e^{\beta|X|}\otimes 1) E_{\lambda}(H_N)\big\| <\infty.
\end{equation}

In the case of only one electron subject to an external potential
$V$ that vanishes at infinity, such as $V_Z$, we saw that
\(\Sigma_{N=1}=\inf\sigma(H_1-V)\). The proper generalization to
$N>1$ is analog to the HVZ theorem for $N$-particle Schr\"odinger
operators. We show that
\begin{equation}\label{hbl:Sigma2}
  \Sigma_N = \min_{N'\geq 1}\{ E_{N-N'}^V+ E_{N'}^0\}
\end{equation}
where $E_{N'}^0$ is the least energy of $N'$ electrons with no
nuclei present, $Z=0$ \cite{Griesemer2002}. Like the HVZ theorem,
\eqref{hbl:HVZ}, for the bottom of the essential spectrum of
$\HamNel{N}$, equation \eqref{hbl:Sigma2} requires the decay of
the interaction between material particles with increasing spacial
separation. While this decay is obvious for the instantaneous
Coulomb interaction, it is more tedious to quantify for the
interaction mediated trough the quantized radiation field. The
main problem in proving \eqref{hbl:Sigma2}, however, is to control
the error which arises when the field energy is split up into two
parts, one associated with the $N'$ electrons far out and one with
the other $N-N'$ electrons. This error is proportional to the
number of photons, as measured by the number operator $N_f$, which
in turn is \emph{not bounded} with respect to the total energy and
thus not under control. To deal with this problem we first proof
\eqref{hbl:Sigma2} with an IR-cutoff $\sigma>0$ in the interaction
and then we show that \eqref{hbl:Sigma2} is obtained in the limit
$\sigma\to 0$ \cite{GLL, Griesemer2002}.

The characterization \eqref{hbl:Sigma2} of the ionization
threshold is important for proving that
$\Sigma_N>\inf\sigma(H_N)$.

\subsubsection{Existence of a ground state}

The dimension of the electron configuration space is inessential
for proving that \eqref{hbl:bind} guarantees existence of a ground
state. Therefore $\inf\sigma(H_N)$ is an eigenvalue of $H_N$
whenever
\begin{equation*}
  \inf\sigma(H_N) < \Sigma_N.
\end{equation*}
However, it is much harder to verify this condition for $N>1$. The
only easy case occurs for spatially confining external potentials
where $\Sigma_N=\infty$. In a tour de force Lieb and Loss recently
showed that
\begin{equation*}
   E_N^Z < \min_{N'\geq 1}\{E_{N-N'}^Z+ E_{N'}^0\}
\end{equation*}
for all atoms and molecules with $Z>N-1$ \cite{LiebLoss2003}.
Combined with \eqref{hbl:Sigma2} this proves that
$E_{N}^Z<\Sigma_N$ and hence that $E_{N}^Z$ is an eigenvalue of
$H_N^Z$ indeed.


\section{Concluding Remarks and Open Problems}
\label{hbl-sec:open}

The results we have described on localization of the electron,
existence of a ground state and existence of scattering states are
established in \cite{Griesemer2002, GLL, FGS1} within the standard
model of QED for non-relativistic electrons (see
Appendix~\ref{sec:qed}). Asymptotic completeness is proved for the
dipole approximation of that model, Hamiltonian \eqref{eq:qed19},
an approximation that is physically reasonable for confined
electrons \cite{FGS2}. We don't expect serious obstacles in
proving ACR for the standard model (with IR cutoff), but to do so
appears prohibitive in view of the additional work due to the
interaction terms quadratic in creation and annihilation
operators. The most important and most interesting open problem in
connection with Rayleigh scattering is to prove completeness
without IR cutoff. This has been done so far only for the
explicitly soluble model of a harmonically bound electron coupled
to radiation in dipole approximation \cite{Arai83}, and for
perturbations thereof \cite{Spohn}. Steps toward ACR for more
general electron Hamiltonians have been undertaken by G\'erard
\cite{Gerard2001}.

The problem of the emitted low energy radiation can be understood
as one aspect of the more general question of the intensity of the
radiation in Rayleigh scattering as a function of the frequency.
Experimentally, sharp spectral lines with frequencies $\omega$
given by Bohr's condition are observed. This condition says that
$\hbar\omega$ is the difference between the energies of two
stationary states, that is, between two eigenvalues of $\Hamel$.
From quantum theory this phenomenon is expected to be a
consequence of the smallness of $\alpha$, which allows one to
compute transition amplitudes in leading order perturbation
theory. Rigorous work in this direction is currently being done by
Bach, Fr\"ohlich and Pizzo \cite{BFP2004}. One also expects that
eigenvalues of $\Hamel$ show up as resonances in the spectrum of
$H_g$ and that the eigenvectors of $H_{g=0}$ are meta-stable
states for the dynamics generated by $H_{g}$, if $g>0$, with a
life-time inversely proportional to the resonance width. Both
these expectations have been confirmed by work of Bach et al, and
by M\"uck \cite{BFS2, Mueck2004}. While the existence of
resonances and meta-stable states is consistent with the
experimentally observed spectral lines, it does not fully account
for them. It remains to be shown that, for small $\alpha$, first
order transitions between meta-stable states dominate the process
of relaxation to the ground state and hence that the intensity is
largest for radiation obeying Bohr's frequency condition. A
related question is the one about a confirmation of the
correspondence principle within QED. By the correspondence
principle, the frequency of radiation emitted by a highly excited
atom agrees with the angular frequency of a classical point charge
on the corresponding Bohr orbit. This principle together with
Bohr's frequency condition determines the distribution of
eigenvalues of highly excited states. A rigorous derivation of the
correspondence principle would therefore confirm -- but not prove
-- the domination of Bohr frequencies at least in the low energy
spectrum.

Many further questions arise once we allow for total energies
\emph{above} the ionization threshold $\Sigma$. Then the atom can
become ionized and the dynamics of the removed electrons is close
to the free one. The first task is thus to study the scattering of
photons at a freely moving electron, the so-called Compton
scattering. This has been done in \cite{FGS3}, where we
established asymptotic completeness for Compton scattering for
energies below a threshold energy that limits the speed of the
electron from above to one-third of the speed of light. To do so,
we had to impose an infrared cutoff, for otherwise no dresses
one-electron states exist.

The natural next step is to combine Rayleigh with Compton
scattering to obtain a complete classification of the long time
asymptotics of matter coupled to radiation. This would include the
photo effect as well as the occurrence of Bremsstrahlung.

There are also very interesting and difficult open questions
related to the binding energy $\Sigma-E_N$ even for $N=1$. From
Section~\ref{hbl-sec:gs} we know that
$$
  E_{N=1}\leq \Sigma + \inf\sigma(-\Delta+V).
$$
That is, if $V\to 0$, the binding energy $\Sigma-E_{N=1}$ with
coupling to the radiation field is at least as large as the
binding energy $-\inf\sigma(-\Delta+V)$ without radiation.
Physical intuition tells us that this binding energy should
actually increase due to the coupling to radiation: the radiation
field accompanying the electron, by the energy-mass equivalence,
adds to the inertia of the electron, that is, makes it heavier and
thus easier to bind. This \emph{mass renormalization} can
explicitly be computed in the dipole approximation and this has
been used to prove enhanced binding by Hiroshima and Spohn
\cite{HiSp2001a}. Without dipole approximation the mass
renormalization is not known explicitly and enhanced binding has
been established so far only for small $\alpha$ \cite{Hainzl2002,
HVV}. It is an interesting and challenging problem to establish
enhanced binding without dipole approximation and for arbitrary
$\alpha$ and $\Lambda$.

Once there are two or more electrons, one would like to know,
first of all, whether two electrons attract or repel each another
in our model of matter. Of course, equal charges repel each other
but this argument neglects the effect of the quantized radiation
field, which is attractive. Two charges close to each other will
share part of their radiation field. Since this reduces the energy
to produce it, binding is encouraged. The questions is thus
whether this binding effect may overcome the Coulomb repulsion.

\appendix
\section{Non-relativistic QED of Atoms and Molecules}
\label{sec:qed}

The purpose of this appendix is to describe atoms and molecules
within UV-regularized, non-relativistic quantum electrodynamics in
Coulomb gauge. We shall also comment on suitable choices of units,
on representations of the theory that avoid the use of
polarization vectors, and on the dipole approximation. For further
information the reader is referred to \cite{BFS1, BraRo2,
CohenTann}.

\subsection{Formal Description of the Model}

To write down the model quickly and in a form familiar from
physics books we shall be somewhat formal at first, using
operator-valued distributions and avoiding domain questions.

The Hilbert space of pure states of $N$ electrons and an arbitrary
number of transversal photons is the tensor product
$\H=\Hel\otimes \F$ where
\begin{equation*}
  \Hel := \wedge_{i=1}^N L^2(\R^3;\C^2),\qquad
  \F := \oplus_{n=0}^{\infty}\otimes_s^n L^2(\R^3;\C^2),
\end{equation*}
$\otimes_s^{n=0}L^2:=\C$, and where $\otimes_s^n L^2(\R^3;\C^2)$,
$n\geq 1$, stands for the symmetrized tensor product of $n$ copies
of $L^2(\R^3;\C^2)$. The vector $\Omega:=(1,0,0,\ldots)\in\F$ is
called \emph{vacuum}. The one-particle wave functions in $\Hel$
and $\F$ are $\C^2$-valued to account for the two spin and the two
polarization states of the electrons and transversal photons,
respectively.

The Hamiltonian of an atom or molecule with static nuclei is a
self-adjoint operator in $\H$ of the form
\begin{equation}\label{eq:qed1}
  H = \sum_{j=1}^N\frac{1}{2m}
  \left[\sigma_j\cdot(-i\nabla_{x_j}+\sqrt{\alpha}A_{\Lambda}(x_j))\right]^2+
  \alpha V_R\otimes 1 + 1\otimes H_f,
\end{equation}
where $\sigma_j=(\sigma_{j,x},\sigma_{j,y},\sigma_{j,z})$ denotes
the triple of Pauli matrices acting on the spin degrees of freedom
of the $j$th electron and $x_j\in \R^3$ is the position of the
$j$th electron. The constant $m>0$ is the (bare) mass of an
electron and $\alpha=e^2/(\hbar c)=e^2$ is the fine structure
constant. In our units $\hbar=1=c$. Another common form of $H$ is
obtained by using that
\begin{equation}\label{eq:qed1b}
  \left[\sigma\cdot(-i\nabla_x+\sqrt{\alpha}A_{\Lambda}(x))\right]^2
  = (-i\nabla_{x}+\sqrt{\alpha}A_{\Lambda}(x))^2 +
  \sqrt{\alpha}\sigma\cdot B(x)
\end{equation}
where \(B(x)=\mathrm{curl} A(x)\).

The operator $V_R$ acts by multiplication with the electrostatic
potential
\begin{equation}\label{eq:qed2}
  V_R(x) = -\sum_{j=1}^K \sum_{i=1}^N \frac{Z_j}{|x_i-R_j|} +
  \sum_{i<j}\frac{1}{|x_i-x_j|}
\end{equation}
of the electrons in the field of $K$ static nuclei with positions
$R_1,\ldots, R_K\in\R^3$ and atomic numbers $Z_1,\ldots, Z_K\in
\Z_{+}$. We use the short-hands $x=(x_1,\ldots,x_N)$ and
$R=(R_1,\ldots,R_N)$.

The operators $H_f$ and $A_{\Lambda}(x)$, for fixed $x\in \R^3$,
are operators on Fock space. $H_f$ has been defined in
Section~\ref{hbl-sec:model} and $A_{\Lambda}(x)$ can be expressed
in the form
\begin{equation}\label{eq:qed3}
  A_{\Lambda}(x)=(2\pi)^{-3/2}
  \sum_{\lambda=1,2}\int_{|k|\leq\Lambda}\frac{d^3k}{\sqrt{2|k|}}
  \left\{\eps_{\lambda}(k)^{*}e^{ik\cdot x}a_{\lambda}(k)+ \eps_{\lambda}(k)
  e^{-ik\cdot x}a_{\lambda}^{*}(k)\right\}.
\end{equation}
The \emph{polarization vectors} $\eps_{\lambda}(k)\in\C^3$,
$\lambda\in\{1,2\}$, are orthogonal to the wave vector $k$ and
normalized
\begin{equation}\label{eq:qed4}
  \eps_{\lambda}^*(k)\cdot \eps_{\mu}(k) =
  \delta_{\lambda\mu}, \qquad \eps_{\lambda}(k)\cdot k=0.
\end{equation}
In addition we assume that $\eps_{\lambda}(tk)=\eps_{\lambda}(k)$
for all $t>0$. The operators $a_{\lambda}^*(k)$ and
$a_{\lambda}(k)$ are \emph{creation- and annihilation operators}
in $\F$. These are operator-valued distributions, formally defined
by $a_{\lambda}(k)\Omega=0$ for all $k\in \R^3$, $\lambda\in
\{1,2\}$, and by the canonical commutation relation
\begin{equation}\label{eq:qed5}
  [a_{\lambda}(k), a_{\mu}^*(q)] =
  \delta_{\lambda\mu}\delta(k-q),\qquad
  [a_{\lambda}^{\#}(k),a_{\mu}^{\#}(q)]=0.
\end{equation}
A rigorous definition of $A_{\Lambda}(x)$ will be given in the
next section. The constant $\Lambda>0$ in \eqref{eq:qed3} is the
\emph{ultraviolet cutoff}. Photons with $|k|>\Lambda$ do not
interact with the electrons under the dynamics generated by $H$.
This is nonphysical but necessary to define $A_{\Lambda,i}(x)$ on
a dense subspace of $\F$. For $\Lambda=\infty$ not even the vacuum
would be in the domain of $A(x)$. In fact, by a formal computation
using the properties of $a_{\lambda}(k)$ and $a_{\lambda}^{*}(k)$,
\(\|A_i(x)\Omega\|^2=\const\int_{|k|\leq \Lambda}|k|^{-1}
d^3k\to\infty\) as $\Lambda\to\infty$.

In the QED of Feynman, Schwinger and Tomanaga, removing the UV
cutoff requires a renormalization of mass, charge and field
strength, a procedure that is mathematically not sufficiently well
understood yet.

\subsection{Atomic Units and Perturbation Theory}

To work in the small-$\alpha$ regime it is convenient to choose
the UV-cutoff $\Lambda$ and the nuclear positions $R_i\in\R^3$
fixed on scales of energy and length where the units are
proportional to the Rydberg energy $m\alpha^2/2=mc^2\alpha^2/2$
and the Bohr radius $1/(m\alpha)=\hbar^2/me^2$. We shall therefore
rewrite the Hamiltonian in these units. It is instructive to begin
by first scaling electron position and photon momentum
independently. Let $U:\H\to\H$ be defined by
\((U\ph)_n(x,k_1,\ldots,k_n)= \eta^{3/2}\mu^{3n/2}\ph_n(\eta x,\mu
k_1,\ldots,\mu k_n)\). Then
\begin{equation}\label{eq:qed6}
\begin{aligned}
  \mu^{-1} UHU^{*} = &\sum_{j=1}^N\frac{1}{2m \eta^2\mu}
  \left[\sigma_j\cdot(-i\nabla_{x_j}+\sqrt{\alpha}\eta\mu
  A_{\Lambda/\mu}(\eta \mu x_j))\right]^2\\ &+
  \frac{\alpha}{\eta\mu} V_{R/\eta}\otimes 1 + 1\otimes H_f,
\end{aligned}
\end{equation}
which is most easily verified using the definition of
$A_{\Lambda}(x)$ given in the next section. In order that
\(2m\eta^2\mu=1\) and \(\eta\mu=\alpha\) we choose
$\eta=(2m\alpha)^{-1}$ and $\mu=2m\alpha^2$. Next we express the
UV cutoff and the nuclear positions in these units, that is we
replace
\begin{equation}\label{eq:qed6b}
  \Lambda/\mu\to \Lambda,\qquad R/\eta\to R,
\end{equation}
a non-unitary change of the Hamiltonian! Thus in the new units the
Hamiltonian reads
\begin{equation}\label{eq:qed7}
  \sum_{i=1}^N
  \left[\sigma_i\cdot(-i\nabla_i+ \alpha^{3/2}A_{\Lambda}(\alpha x_i))\right]^2+
  V_R\otimes 1 + 1\otimes H_f
\end{equation}
where the dependence on $\alpha$ is concentrated in
electron-photon interaction $\alpha^{3/2}A_{\Lambda}(\alpha x)$.
The papers by Bach et al. concern the Hamiltonian \eqref{eq:qed7},
many others concern \eqref{eq:qed1}. When comparing results that
are valid for small $\alpha$ only, one must keep in mind that
these Hamiltonians are not equivalent, not even for atoms: the
substitution $\Lambda\to \Lambda\alpha^2$, which occurs in
\eqref{eq:qed6b}, corresponds to the change $m\mapsto m/\alpha^2$
of the electron mass, as follows from \eqref{eq:qed6} with
$\eta=\mu^{-1}=\alpha^2$.

\subsection{Fock-Spaces, Creation- and Annihilation Operators}

We next give a rigorous definition of the quantized vector
potential $A_{\Lambda}(x)$ and we shall comment on the
self-adjointness of $H$. In order to prepare the ground for the
next section we define Fock space, creation- and annihilation
operators in larger generality then needed here. A good reference
for this section is \cite{BraRo2}.

Given a complex Hilbert space $\h$ the bosonic Fock space over
$\h$,
\begin{equation}\label{eq:qed8}
     \F=\F(\h) = \oplus_{n\geq 0} \mathcal{S}_n(\otimes^n\h)
\end{equation}
is the space of sequences $\ph=(\ph_n)_{n\geq 0}$, with $\ph_0\in
\C$, $\ph_n\in \mathcal{S}_n(\otimes^n\h)$, and $\sum_{n\geq
0}\|\ph_n\|^2<\infty$. Here $\mathcal{S}_n$ denotes the orthogonal
projection onto the subspace of symmetrized tensor products of $n$
vectors in $\h$. The inner product in $\F$ is defined by
\begin{equation*}
  \sprod{\ph}{\psi} = \sum_{n\geq 0}(\ph_n,\psi_n),
\end{equation*}
where $(\ph_n,\psi_n)$ denotes the inner product in $\otimes^n
\h$. We use $\Fin$ to denote the dense subspace of vectors
$\ph\in\F$ with $\ph_n=0$ for all but finitely many $n\in\N$.

Given $h\in\h$ the creation operator $a^{*}(h):\Fin\subset\F\to
\F$ is defined by
\begin{equation}\label{eq:qed9}
  [a^{*}(h)\ph]_n = \sqrt{n}\mathcal{S}_n(h\otimes \ph_{n-1})
\end{equation}
and the annihilation operator $a(h):\Fin\subset\F\to \F$ is the
restricted to $\Fin$ of the adjoint of $a^{*}(h)$. The operators
$a(h)$ and $a^{*}(h)$ satisfy the canonical commutation relations
(CCR)
\begin{equation*}
[a(g),a^{*}(h)] = (g,h),\qquad [a^{\#}(g),a^{\#}(h)] = 0.
\end{equation*}
In particular, $[a(h),a^{*}(h)]=\|h\|^2$ which implies that
$\|a(h)\ph\|+\|\ph\|$ and $\|a^{*}(h)\ph\|+\|\ph\|$ are equivalent
norms. It follows that the closures of $a(h)$ and $a^{*}(h)$ have
the same domain. On this domain $a^{*}(h)$ is the adjoint of
$a(h)$ \cite[Theorem 5.2.12]{BraRo2}. The operator
\begin{equation}\label{eq:qed10}
  \phi(h) = \frac{1}{\sqrt{2}}(a(h)+a^{*}(h))
\end{equation}
is essentially self-adjoint on $\Fin$ \cite{BraRo2}. It is useful
to note that
\begin{equation*}
  [\phi(g),\phi(h)] = i\Ima(g,h).
\end{equation*}

In the case of QED, $\h= L^2(\R^3;\C^2)$ with inner product
\((g,h)=\sum_{\lambda=1,2}\int
\overline{g_{\lambda}(k)}h_{\lambda}(k)\, d^3k\) and
$A_i(x)=\phi(G_{x,i})$, $G_{x,i}\in \h$, $i=1,2,3$, being the
components of
\begin{equation}\label{eq:qed11}
  G_x(k,\lambda) =
  \frac{\kappa(k)}{|k|^{1/2}}\eps_{\lambda}(k)e^{-ik\cdot x},
\end{equation}
where $\kappa(k) = (2\pi)^{-3/2}\chi_{|k|\leq\Lambda}(k)$. More
generally we may allow $\kappa$ to be any real-valued, spherically
symmetric function with $\kappa/\sqrt{\omega}\in L^2(\R^3)$. In
particular $|\kappa(-k)|=|\kappa(k)|$ which implies that
$[A_i(x),A_j(y)]=0$ for all $x,y\in \R^3$ and $i,j\in \{1,2,3\}$.

The Hamiltonian~\eqref{eq:qed1} is defined and symmetric on the
dense subspace
\begin{equation}\label{eq:qed11b}
  \mathcal{D} = \big[\wedge_{i=1}^N C_0^{\infty}(\R^3;\C^2)\big]\otimes
  \Fin(C_0^{\infty}(\R^3;\C^2))
\end{equation}
where $\Fin(C_0^{\infty}(\R^3;\C^2))$ is the space of vectors
$\ph=(\ph_n)_{n\geq 0}\in\Fin$ with
\(\ph_n=\otimes^nC_0^{\infty}(\R^3;\C^2)\). Let
$H_0=-\Delta\otimes 1+1\otimes H_f$. Then $H_0$ is essentially
self-adjoint on $\mathcal{D}$, self-adjoint on
$D(H_0)=D(-\Delta\otimes 1)\cap D(1\otimes H_f)$, and $H-H_0$ is
bounded relative to $H_0$. It follows that the closure of
$H\restricted\mathcal{D}$ is defined on $D(H_0)$ and symmetric on
this domain. Since $H$ is self-adjoint on $D(H_0)$, according to
Hiroshima \cite{Hiroshima2002}, we conclude that $H$ is
essentially self-adjoint on $\mathcal{D}$. Alternatively, the
Hamiltonian \eqref{eq:qed1} may be self-adjointly realized in
terms of the Friedrichs' extension of $H\restricted\mathcal{D}$,
since this operator is bounded from below, or, by using the
theorem of Kato-Rellich for $\Lambda/\alpha$ small enough
\cite{BFS2}. ($\alpha\Lambda$ small enough for the Hamiltonian
\eqref{eq:qed7}.)

\subsection{Avoiding Polarization Vectors}

The fact that the polarization vectors are necessarily
discontinuous as functions of $\hat{k}=k/|k|\in S^2$, by a
well-known result of H.~Hopf, may lead to annoying technical
problems \cite{GLL}. To show how these problems can be avoided we
construct a representation of $H$ that does not depend on a choice
of polarization vectors \cite{FGS2, LiebLoss2004}. This
representation is based on a description of single-photon states
by vectors in
\begin{equation}\label{eq:qed12}
  \h_{T}:= \{h\in L^2(\R^3;\C^3)| h(k)\cdot k = 0\quad \text{for all}\quad
  k\},
\end{equation}
the space of transversal photons. Let $u: L^2(\R^3;\C^2)\to \h_T$
be the unitary map
\begin{equation}\label{eq:qed13}
  u:(h_1,h_2) \mapsto \sum_{\lambda=1,2}
  h_{\lambda}\eps_{\lambda}^{*}
\end{equation}
where $\{\eps_{\lambda}\}_{\lambda=1,2}$ are the polarization
vectors employed in the definition of $H$, an let $U:\F\to
\F(\h_{T})$ be defined by
\begin{equation*}
  Ua^{*}(h)U^{*}=a^*(uh), \qquad U\Omega=\Omega.
\end{equation*}
It follows that $UA_i(x)U^{*}=\phi(uG_{x,i})$ where
\begin{align}
  (uG_{x,j})(k) &= \sum_{\lambda=1,2}\frac{\kappa(k)}{|k|^{1/2}}
  \eps_{\lambda}(k)_j \eps_{\lambda}(k)^{*}\nonumber \\
  &= \frac{\kappa(k)}{|k|^{1/2}}e^{-ik\cdot x} (e_j - \hat{k}(\hat{k}\cdot
  e_j)), \label{eq:qed14}
\end{align}
$\{e_1,e_2,e_3\}$ being the canonical basis of $\R^3$. Note that
$\phi(uG_{x,j})$ is \emph{one} operator and not a triple of
operators even though $k\mapsto uG_{x,j}(k)$ is a vector-valued
function.

The Hamilton operator $H_T=UHU^*$ is the desired new
representation of $H$. It has the form of $H$ in \eqref{eq:qed1}
with the only difference that the form-factor of $A(x)$ is now
given by \eqref{eq:qed14}, a function in
$C^{\infty}(\R^3\backslash\{0\})$.

By choosing other unitary mappings $u$ from $L^2(\R^3;\C^2)$ onto
$\h_{T}$ one may define more equivalent representations of QED.
For example the map
\begin{equation}\label{eq:qed15}
  u_{2}: (h_1,h_2) \mapsto \sum_{\lambda=1,2}
  h_{\lambda}\eps_{\lambda}^{*} \wedge \hat{k}
\end{equation}
leads to the representation of $H$ where the quantized vector
potential is defined in terms of the form factor
\begin{equation}\label{eq:qed15b}
  (u_{2}G_{x,i})(k) = \frac{\kappa(k)}{|k|^{1/2}}e^{-ik\cdot x}
(e_i\wedge \hat{k}),
\end{equation}
the choice preferred in \cite{LiebLoss2004}.

For the mathematical analysis of systems of electrons interacting
with photons it is often necessary to localized the photons in
their position space. That is, the photon wave function $h\in\h_T$
is mapped to $J(i\nabla_k)h$ where $J\in C_0^{\infty}(\R^3)$. Now
$J(i\nabla_k)h\not\in \h_{T}$ unless $J=0$ or $h=0$, and
projecting $J(i\nabla_k)h$ back to $\h_T$ would destroy the
localization accomplished by the operator $J(i\nabla_k)$. The
solution to this problem is to work on the enlarged one-boson
Hilbert space $\h_{\rm ext}=L^2(\R^3;\C^3)=\h_{T}\oplus \h_{L}$
which also includes the space of longitudinal photons
\(\h_{L}=\{h|k\wedge h(k)=0\}\). The Hilbert space for the entire
system becomes $\H_{\rm ext}=\Hel\otimes \F(\h)\simeq \H\otimes
\F(\h_{L})$ and we define a Hamiltonian on $\H_{\rm ext}$ by
\begin{align*}
H_{\rm ext} &= H_T\otimes 1 + 1\otimes H_{f,L}\\
        &= \sum_{j=1}^{N}\frac{1}{2m}[\sigma_{j}\cdot(-i\nabla_{x_j}+\alpha^{1/2}A(x))]^2
        + \alpha V_R + H_f
\end{align*}
where $A_j(x)=\phi(uG_{x,j})$ as above, but now $uG_{x,j}$ is
considered as an element of $\h_{\rm ext}=L^2(\R^3;\C^3)$. The
fake longitudinal bosons from $\h_{L}$ do not interact with the
electrons and hence do not affect the dynamical properties of the
system. However, by definition of $H_{\rm ext}$ they contribute
additively to the total energy and need to be projected out at the
end of any analysis of the energy spectrum.

To conclude this section we return to a more formal representation
of $A(x)$ by expanding photon wave functions in terms of
$\delta$-distributions. Let $\delta_k(q)=\delta(q-k)$. We define,
formally,
\begin{equation*}
  a_j^{\#}(k) = a^{\#}(e_j\delta_k),\qquad
  a^{\#}(k)=(a_1^{\#}(k),a_2^{\#}(k),a_3^{\#}(k)).
\end{equation*}
From the expansion
  $h(k) = \sum_{j=1}^3\int h_j(q)e_j\delta_{k}(q)\, d^3q$
and the (semi)-linearity of $a^{\#}(h)$ we obtain
\begin{align*}
  a(h) &= \sum_{j=1}^3 \int \overline{h_j(k)}a_j(k)\, d^3k =
  \int\overline{h(k)}\cdot a(k)\, d^3k\\
  a^{*}(h) &= \sum_{j=1}^3 \int h_j(k)a_j^{*}(k)\, d^3k =
  \int h(k)\cdot a^{*}(k)\, d^3k.
\end{align*}
In particular, in the representation defined by \eqref{eq:qed14},
\begin{equation*}
  A(x) = \int \frac{\kappa(k)}{|k|^{1/2}} P(k)\big\{e^{ik\cdot x}
  a(k) + e^{-ik\cdot x}a^{*}(k)\big\} \, d^3k
\end{equation*}
where $P(k)$ denotes the orthogonal projection onto the plane
perpendicular to $k$. If \eqref{eq:qed15b} is used then
\begin{equation*}
  A(x) = \int \frac{\kappa(k)}{|k|^{1/2}}
  \hat{k}\wedge\big\{e^{ik\cdot x}
  a(k) + e^{-ik\cdot x}a^{*}(k)\big\}\, d^3k.
\end{equation*}

\subsection{The Dipole Approximation}

In the dipole approximation of QED the quantized vector potential
$A(x)$ in the Hamilton \eqref{eq:qed1} is replaced by $A(0)$. By
\eqref{eq:qed1b} the Hamiltonian \eqref{eq:qed1} then reduces to
\begin{equation}\label{eq:qed16}
  H_{\rm dip}= \sum_{j=1}^N \frac{1}{2m}(-i\nabla_{x_j}+\alpha^{1/2}A(0))^2 + \alpha V_{R} +H_f
\end{equation}
where the interaction with the electron spin has dropped out.
Without loss of generality we may now describe the electrons by
vectors in the smaller space $\Hel=\wedge_{i=1}^N L^2(\R^3)$ of
spin-less $N$-fermion systems.

The ``constant'' vector potential in \eqref{eq:qed16} may be
gauged away with the help of the operator-valued gauge
transformation
\begin{equation}\label{eq:qed17}
  U = \exp\left(\alpha^{1/2}\sum_{i=1}^N x_i\cdot A(0)\right),
\end{equation}
also known as Pauli-Fierz transformation. Since
$U(-i\nabla_{x_j})U^*=-i\nabla_{x_j}-\alpha^{1/2}A(0)$,
$UA(0)U^{*}=A(0)$, and
\begin{equation*}
  UH_fU^* = H_f + \sqrt{\alpha}\sum_{j=1}^N x_j\cdot E(0) + \alpha \|\kappa\|^2\Big(\sum_{j=1}^n
  x_j\Big)^2
\end{equation*}
where \(E(0)=-i[H_f,A(0)]\) is the quantized electric field, we
arrive at
\begin{equation}\label{eq:qed18}
\begin{aligned}
  UH_{\rm dip}U^*= &\sum_{j=1}^N \left(-\frac{1}{2m}\Delta_{x_j}+\sqrt{\alpha}x_j\cdot
  E(0)\right)\\ &+ \alpha V_R + H_f+ \alpha\|\kappa\|^2\Big(\sum_{j=1}^N x_j\Big)^2.
\end{aligned}
\end{equation}
The dipole approximation seems justified when all electrons are
localized in a small neighborhood of the origin $x=0$, that is,
when the total energy is below the ionization threshold. It then
seems equally justified to drop the last term in \eqref{eq:qed18}
and to multiply $x\cdot E(0)$ with a space cutoff $g\in
C_0^{\infty}(\R^3)$; the later serves to ensure that the
Hamiltonian $H$ remains semi-bounded (after dropping the last
term). This leads us to
\begin{equation}\label{eq:qed19}
  \tilde{H}_{\rm dip} = \sum_{j=1}^N \left(-\frac{1}{2m}\Delta_{x_j}+\sqrt{\alpha}g(x_j)x_j\cdot
  E(0)\right)+ \alpha V_R + H_f,
\end{equation}
which is also called dipole approximation of \eqref{eq:qed1}. It
has the advantage, over \eqref{eq:qed16}, to be linear in creation
and annihilation operators, which may simplify the analysis.

The Pauli-Fierz transformation \eqref{eq:qed17} is very useful in
the analysis of the original Hamiltonian \eqref{eq:qed1} as well.
Its effect is to replace $A(x)$ by $A(x)-A(0)=\phi(G_x-G_0)$ where
\begin{equation*}
  |G_x(k)-G_0(k)| = \left|\frac{\kappa(k)}{\sqrt{|k|}}(e^{ik\cdot
  x}-1)\right|\leq |k|^{1/2}|\kappa(k)||x|.
\end{equation*}
Thus the IR-singularity of the form-factor in $UHU^*$ is reduced
by one power of $|k|$ at the expense of the unbounded factor
$|x|$. This factor, however, is compensated by the exponential
decay whenever the total energy is below the ionization threshold
(see Section~\ref{hbl-sec:expo}).


\end{document}